\documentclass{emulateapj}

\slugcomment{Draft version \today}

\shorttitle{Typing SNRs Using X-ray Line Morphologies}
\shortauthors{LOPEZ ET AL.}

\begin{document}

\title{Typing Supernova Remnants Using X-ray Line Emission Morphologies}

\author{L.~A. Lopez\altaffilmark{1,6}, E.~Ramirez-Ruiz\altaffilmark{1}, C. Badenes\altaffilmark{2}, D.~Huppenkothen\altaffilmark{3},
  T. E. Jeltema\altaffilmark{4}, D. A. Pooley\altaffilmark{5} }
%%%
\altaffiltext{1}{Department of Astronomy and Astrophysics, University of California Santa Cruz, 159 Interdisciplinary Sciences Building,
  1156 High Street, Santa Cruz, CA 95064, USA; lopez@astro.ucsc.edu.}
\altaffiltext{2}{Department of Astrophysical Sciences, Princeton University Peyton Hall, Ivy Lane, Princeton, NJ 08544, USA}
\altaffiltext{3}{Astronomical Institute, "Anton Pannekoek", University of Amsterdam, PO Box 94249, 1090 GE Amsterdam, the Netherlands}
\altaffiltext{4}{UCO/Lick Observatories, 1156 High Street, Santa Cruz, CA, 95064, USA} 
\altaffiltext{5}{Astronomy Department, University of  Wisconsin, 475 North Charter St, Madison, WI 53706, USA}
\altaffiltext{6}{NSF Graduate Research Fellow}

\begin{abstract}
We present a new observational method to type the explosions of young supernova remnants (SNRs). By measuring the morphology of the {\it Chandra} X-ray line emission in seventeen Galactic and Large Magellanic Cloud SNRs with a multipole expansion analysis (using power ratios), we find that the core-collapse SNRs are statistically more asymmetric than the Type Ia SNRs. We show that the two classes of supernovae can be separated naturally using this technique because X-ray line morphologies reflect the distinct explosion mechanisms and structure of the circumstellar material. These findings are consistent with recent spectropolarimetry results showing that core-collapse SNe are intrinsically more asymmetric.

\end{abstract}


\section{Introduction}

Over the past few decades, observations and theory of supernovae explosions (SNe) and their remnants have advanced tremendously. The original classification of SNe was based on their optical spectra near maximum brightness (Minkowski 1941; see Filippenko 1997 for a review), when features originate from the outermost layers of the exploding stars. At later times, SN expansion causes the chemical and physical structure of the SN ejecta and the circumstellar material (CSM) to dominate the spectra. Study of these nebular spectra revealed two physically disinct classes of SNe: Type Ia and core-collapse (CC; Types Ib, Ic, and II) explosions. There is a broad consensus now that Type Ia SNe constitute the thermonuclear detonation of a white dwarf, while CC SNe follow the gravitational collapse of a massive ($>8 M_{\sun}$) star. These different pathways produce distinct observational signatures that reflect their underlying explosion mechanisms. 

As the ejecta continue to expand following an explosion, they are heated to X-ray emitting temperatures by collisionless shocks. Due to this phenomenon, X-ray emission lines are observable in young supernova remnants (SNRs with ages $<$10,000 years) from the metals produced within the stellar interiors and during the SNe. Thus, study of X-ray spectra and X-ray morphology can reveal much about SNe progenitors (see Weisskopf \& Hughes 2006 for a review). For example, the strength of X-ray emission lines can be used to infer chemical abundances and to compare with the values predicted by Type Ia and CC SNe models. In particular, the O/Fe ratio is often used for this purpose, since Type Ia SNe produce comparatively more Fe than CC SNe, and CC SNe yield large amounts of O. However, the analysis of ejecta-dominated X-ray spectra is very challenging given our limited understanding of a number of topics, including the hydrodynamic interaction between the ejecta and the ambient medium, the inner structure of the ejecta, and the electron heating processes at the reverse shock (Badenes et al. 2003, Rakowski et al. 2006). So far, this detailed analysis of the ejecta emission including all these subtleties has only been done for a few objects.

In this Letter, we present a new, observational method for the typing of young SNRs. Specifically, we describe how the asymmetry of \hbox{X-ray} line emission from SNRs can be used to discern between a Type Ia or a CC origin. Our technique is general and quantitative, and it is independent of the conditions of the plasma. Thus, this approach can be used systematically for comparison between many sources.

\begin{deluxetable*}{cccccccc}[ht!]
\tablecolumns{8} \tablewidth{0pc} \tabletypesize{\footnotesize}
\setlength{\tabcolsep}{0.0in} \renewcommand{\arraystretch}{1.0}
\tablewidth{0pt} \tablecaption{Archival SNRs, Sorted by Age}
\tablehead{\colhead{Source} & \colhead{ObsID} & \colhead{ACIS Exp.} & \colhead{Age} & \colhead{Distance} & \colhead{Radius\tablenotemark{a}} & 
  \colhead{$L_{X}$\tablenotemark{b}} & \colhead{References}
  \\ \colhead{} & \colhead{} & \colhead{(ks)} & \colhead{(years)} &
  \colhead{(kpc)} & \colhead{(pc)} & \colhead{($\times 10^{37}$ erg/s)} & \colhead{}}
\startdata 
\cutinhead{Type Ia Sources} 
0509$-$67.5 & 776, 7635, 8554 & 113 & 350--450 & 50 & 5.96 & 1.76 & \cite{bad08} \\
Kepler & 6714--6718, 7366 & 751 & 405 & 5.0 & 3.88 & 0.19 & \cite{reynolds} \\
Tycho & 3887 &  150 & 437 & 2.4 & 3.72 & 0.12 & \cite{warren05} \\
0519$-$69.0 & 118 & 40 & 400--800 & 50 & 6.56 & 1.06 & \cite{rest05} \\
N103B & 125 & 37 & 860 & 50 & 5.96 & 1.67 & \cite{lewis03} \\
DEM L71 & 775, 3876, 4440 & 148 & $\sim$4360 & 50 & 11.9 & 0.77 & \cite{hughes03} \\ 
0548$-$70.4 & 1992 & 60 & $\sim$7100 & 50 & 17.9 & 0.96 & \cite{hendrick} \\
\cutinhead{Core-collapse Sources}
Cas A & 4634--4639, 5196, 5319--5320 & 993 & 309--347 & 3.4 & 3.77 & 2.58 & \cite{hwang04} \\
W49B & 117 & 55 & $\sim$1000 & 8.0 & 6.30 & 4.48 & \cite{lal} \\
G15.9$+$0.2 & 5530, 6288, 6289 & 30 & $\sim$1000 & 8.5 & 8.72 & 1.38 & \cite{reynolds06} \\
G11.2$-$0.3 & 780--781, 2322, 3909-3912 & 95 & 1623 & 5.0 & 4.17 & 1.18 & \cite{kaspi01} \\
Kes 73 & 729 & 30 & 500--2200 & 8.0 & 6.11 & 0.94 & \cite{gotthelf} \\
RCW 103 & 970 & 49 & $\sim$2000 & 3.3 & 5.43 & 2.21 & \cite{carter} \\
N132D & 5532, 7259, 7266 & 90 & $\sim$3150 & 50 & 21.47 & 9.92 & \cite{bork07} \\
G292.0$+$1.8 & 6677--6680, 8221, 8447 & 516 & $\sim$3300 & 6.0 & 9.45 & 0.56 & \cite{park07} \\ 
N49B & 1041 & 35 & 10000 & 50 & 23.85 & 2.39 & \cite{park03} \\
B0453$-$685 & 1990 & 40 & 13000 & 50 & 20.27 & 0.54 & \cite{gaensler03} \\
\enddata 
\tablenotetext{a}{Radius $R$ used in Eq. 3 of the power-ratio calculation. This radius is selected to enclose the entire source in the full-band X-ray image, and it is determined assuming the distances listed above.}  
\tablenotetext{b}{X-ray luminosity in the 0.5--2.1 keV band, from the {\it Chandra} SNR Catalog. The value for G15.9$+$0.2 is calculated using an absorbed planar shock (vpshock) model of the integrated spectra \citep{reynolds06}.}
\end{deluxetable*}

\section{Observations and Methods}

\subsection{Sample and Data Analysis}

We analyzed archival {\it Chandra} ACIS observations of the seventeen SNRs listed in Table 1. Seven of our sources are thought to be Type Ia SNe, while ten are considered to have originated from a CC SNe (see Table 1). We selected only sources in the Milky Way and LMC that had their entire X-ray emitting area observed with one ACIS pointing; this criterion excluded sources with large spatial extent, such as SN 1006. We also required the targets to have strong line emission attributed to the ejecta, with Si {\sc xiii} line emission surface brightnesses $>$0.01 counts/pixel$^2$ within the apertures $R$ listed in Table 1. This criterion excluded SNRs with weak lines and those dominated by continuum emission (bremsstrahlung or non-thermal), e.g. G1.9$+$0.3. This requirement removed two SNRs whose X-ray line emission is attributed to shocked interstellar medium (ISM) and not ejecta: SNR 0506$-$68.0 \citep{hughes06} and 3C 391 \citep{chen04}. Additionally, we excluded sources with large-scale morphologies that are the result of axisymmetric winds from pulsars (like Crab or 3C 58). One other source, N63A, was removed since its morphology is distorted by its association with an HII region and obscuration by an intervening dense cloud \citep{chu88,warren03}. The final sample reflects all SNRs that meet the above criteria.

Each source was observed for $\sim$30--1000 ks. Data reduction and analysis was performed using the {\it Chandra} Interactive Analysis of Observations ({\sc ciao}) Version 4.0. We followed the {\sc ciao} data preparation thread to reprocess the Level 2 X-ray data, and we extracted the global {\it Chandra} X-ray spectra of each source using the {\sc ciao} command {\it specextract}. Exposure-corrected images of the Si {\sc xiii} line were then produced for each source by filtering data to the narrow energy range of that feature ($\sim$1.75--2.00 keV). The Si {\sc xiii} line was selected for our comparative analysis as it is the only strong emission line common to all seventeen sources. For sources with bright pulsars (G11.2$-$0.3, Kes 73, RCW 103, G292.0$+$1.8), the pulsar location and extent was identified using the {\sc ciao} command {\it wavdetect} (a source detection algorithm using wavelet analysis; Freeman et al. 2002). We replaced the region identified by {\it wavdetect} with pixel count values selected from the Poisson distribution of the area surrounding the pulsar using the {\sc ciao} command {\it dmfilth}. This process removed the bright pulsar while preserving the morphology of the emission surrounding the pulsar. No other point sources were removed because of potential confusion with small ejecta clumps. 

Given the young-to-middle age of our sources (see Table 1), we expect that all are ejecta-dominated. X-ray spectral modeling generally confirms that abundances are above those of the ISM (see references in Table 1), suggesting the emission is indeed dominated by the shocked SN ejecta and not by shocked ISM. In the case of G292.0$+$1.8, \cite{park04} show that the Si {\sc xiii} and S {\sc xv} is contaminated by shocked ISM in its ``belt''. We note that our analysis of Ne {\sc ix} and Mg {\sc xi} in this source (which Park et al. 2004 attribute to ejecta) gave results consistent those of the Si {\sc xiii} line. 

\subsection{Methods}

To measure the asymmetry of the X-ray line emission, we have applied a power-ratio method (PRM) to the Si {\sc xiii} images of our seventeen sources. The PRM was first applied to quantify the X-ray morphology of galaxy clusters observed with {\it  ROSAT} (Buote \& Tsai, 1995, 1996) and with {\it Chandra} \citep{j05}. In \cite{lal}, we employed the method to compare the symmetry and elongation of line emission in the complex SNR W49B. We refer the reader to these papers for a detailed formalism; here, we give a brief overview of the technique.

\begin{figure*}
\includegraphics[width=0.45\textwidth,angle=-90]{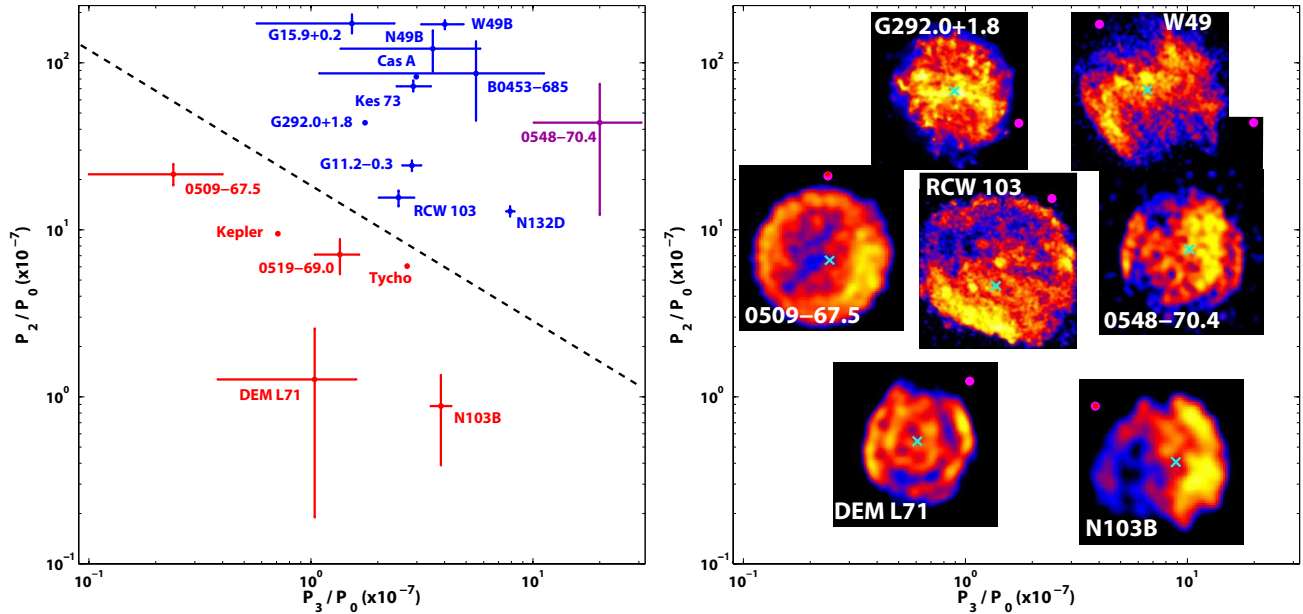}
\caption{Left: $P_{2}/P_{0}$ versus $P_{3}/P_{0}$ for sources classified as core-collapse SNRs (blue) and the Type Ia SNRs (red). SNR 0548$-$70.4 is plotted in purple due to its anomalous ejecta properties (see Section 3). Right: Si {\sc xiii} images of seven sources, plotted in the location of their power ratio values. Cyan Xs mark their full-band centroids, and purple dots are placed at their locations on the power-ratio plot. The images of RCW 103 and G292.0$+$1.8 have their pulsars removed using the method outlined in $\S$2.}
\label{fig:p2p3}
\end{figure*}

The PRM measures asymmetries in an image via calculation of the multipole moments of the X-ray surface brightness in a circular aperture. It is derived similarly to the multipole expansion of the two-dimensional gravitational potential within an enclosed radius $R$:

\begin{eqnarray}
\lefteqn{\Psi(R,\phi) = -2Ga_0\ln\left({1 \over R}\right)-2G }
\nonumber \\ & & \times \sum^{\infty}_{m=1} {1\over m
  R^m}\left(a_m\cos m\phi + b_m\sin
m\phi\right), \label{eqn.multipole}
\end{eqnarray}

\noindent
where the moments $a_m$ and $b_m$ are
\begin{eqnarray}
a_m(R) & = & \int_{R^{\prime}\le R} \Sigma(\vec x^{\prime})
\left(R^{\prime}\right)^m \cos m\phi^{\prime} d^2x^{\prime}, \nonumber \\
b_m(R) & = & \int_{R^{\prime}\le R} \Sigma(\vec x^{\prime})
\left(R^{\prime}\right)^m \sin m\phi^{\prime} d^2x^{\prime}, \nonumber
\end{eqnarray}

\noindent
$\vec x^{\prime} = (R^{\prime},\phi^{\prime})$, and $\Sigma$ is the surface mass density. For our imaging analyses, the X-ray surface brightness replaces surface mass density in the power ratio calculation.

The powers of the multipole expansion are obtained by integrating the magnitude of $\Psi_m$ (the \textit{m}th term in the multipole expansion of the potential) over a circle of radius $R$,

\begin{equation}
P_m(R)={1 \over 2\pi}\int^{2\pi}_0\Psi_m(R, \phi)\Psi_m(R, \phi)d\phi.
\end{equation}

\noindent
Ignoring the factor of $2G$, this equation reduces to

\begin{eqnarray}
P_0 & = & \left[a_0\ln\left(R\right)\right]^2 \nonumber \\
P_m & = & {1\over 2m^2 R^{2m}}\left( a^2_m + b^2_m\right) 
\end{eqnarray}

The moments $a_{m}$ and $b_{m}$ (and consequently, the powers $P_{m}$)
are sensitive to the morphology of the X-ray surface brightness
distribution, and higher-order terms measure asymmetries at
successively smaller scales relative to the position of the aperture
center (the origin). To normalize with respect to flux, we divide the
powers by $P_{0}$ to form the power ratios, $P_m/P_0$. $P_{1}$
approaches zero when the origin is placed at the surface-brightness
centroid of an image, so we have set the aperture center in all
analyses to the full-band (0.5--8.0 keV) centroid of each remnant. In
this case, morphological information is given by the higher-order
terms. $P_{2}/P_{0}$ is the quadrupole ratio; examples of sources that
give high $P_{2}/P_{0}$ are those with elliptical morphologies and
those with off-center centroids because one side is substantially brighter than the other. $P_{3}/P_{0}$ is the octupole ratio; examples of sources that give high $P_{3}/P_{0}$ are those with deviations from mirror symmetry relative to their centroid.

A Monte Carlo approach described in \cite{lal} is used to estimate the uncertainty in the power ratios. Specifically, the exposure-corrected images (normalized to have units of counts) are adaptively-binned using the program {\it AdaptiveBin} \citep{s01} such that all zero pixels are removed to smooth out noise. Then, noise is added back in by taking each pixel intensity as the mean of a Poisson distribution and selecting randomly a new intensity from that distribution. This process was repeated 100 times for each Si {\sc xiii} image, creating 100 mock images per source. The 1-$\sigma$ confidence limits represent the sixteenth highest and lowest power ratio obtained from the 100 mock images of each source.

\section{Discussion}

The calculated power ratios $P_2/P_0$ and $P_3/P_0$ for the Si {\sc xiii} images of our sources are plotted in Figure~\ref{fig:p2p3} (left), with example Si {\sc xiii} images of seven sources shown in Figure~\ref{fig:p2p3} (right). The SNRs with elongated or barrel-like features (like W49B, Cas A, and G292.0$+$1.8) have the largest $P_{2}/P_{0}$ values, and sources that are more compact and circular (e.g., 0519$-$69.0 and DEM L71) have the lowest $P_{2}/P_{0}$. Some sources that appear spherical by eye (like 0509$-$67.5 or 0548$-$70.4) have large $P_{2}/P_{0}$ because they are brighter on one side and their full-band centroids have off-center positions. SNRs with obvious mirror asymmetries (such as N132D and N103B) produce large $P_{3}/P_{0}$, while sources with line emission distributed evenly around their full-band centroids (e.g., DEM L71 and 0509$-$67.5) have small \hbox{$P_{3}/P_{0}$}.

Overall, we find that the power ratios of the Type Ia SNRs are significantly different than those of the CC SNRs. The CC SNe have roughly an order of magnitude larger quadrupole power ratio: the mean $P_{2}/P_{0}$ of the Type Ia SNe is (7.7$\pm$0.7)$\times 10^{-7}$ with a standard deviation of 6.9$\times10^{-7}$ (excluding SNR 0548$-$70.4, see discussion below) and of the CC SNe is (80.3$\pm$2.6)$\times 10^{-7}$ with a standard deviation of 56.1$\times10^{-7}$. This result implies that the CC SNR line emission is comparatively much more elliptical or have more off-center centroids than Type Ia SNRs. We find that the mean $P_{3}/P_{0}$ are also different: the mean of the Type Ia SNe is (1.7$\pm$0.1)$\times 10^{-7}$ (excluding SNR 0548$-$70.4) with a standard deviation of 2.1$\times10^{-7}$ and of the CC SNe is (3.5$\pm$0.3)$\times 10^{-7}$ with a standard deviation of 1.8$\times10^{-7}$. This result indicates that the X-ray line emission in CC SNe is more mirror asymmetric than Type Ia SNe. Additionally, we find that the Type Ia SNRs with greater ellipticity or off-center centroids (larger $P_{2}/P_{0}$) tend to have larger mirror symmetry (smaller $P_{3}/P_{0}$), and those with less mirror symmetry are significantly more circular or balanced surface brightness (smaller $P_{2}/P_{0}$). Overall, the Type Ia and CC SNe seem to naturally form two distinct populations in the $P_2/P_0$ -- $P_3/P_0$ plane.

Only one source of the seventeen is an outlier in Figure \ref{fig:p2p3}, SNR 0548$-$70.4. This source has two bright limbs as well as a large, luminous area that is off-center and possibly seen in projection (see the right panel of Figure \ref{fig:p2p3}). \cite{hendrick} categorized this source as a Type Ia based on its low O/Fe ratio of $\sim$16, an intermediate value between the O/Fe ratio of $\sim$0.75 expected for a Type Ia \citep{iwa} and the O/Fe ratio of 70 predicted for a 20 $M_{\sun}$ CC SNe; these authors present several possible explanations for an elevated oxygen abundance in a Type Ia SN. In our opinion, further study of SNR 0548$-$70.4 is warranted to explore its explosion type and anomalous ejecta properties. 

By including targets from the LMC and the Milky Way, we aimed to limit any biases in our sample. Our SNe span a large range of ages and X-ray luminosities (see Table 1). On average, our Type Ia sources have younger estimated ages than the CC SNe, but we find no trends in the power ratios with age or with radius ($R$ from Table 1). The 0.5--2.1 keV X-ray luminosities of the Type Ia and CC SNe are approximately similar, so our analysis is not biased by the brightness of the sources. We note here that we have performed the same analyses for other X-ray emission lines (e.g., Ne {\sc ix}, Mg {\sc xi}, S {\sc xv}), and all produce the the two distinct populations in the $P_2/P_0$ -- $P_3/P_0$ plane.

Our results show that as a class, CC SNRs are much more asymmetric or elliptical than Type Ia SNRs. These morphological differences reflect the distinct explosions and CSM structures of the progenitors of Type Ia and CC SNe. This finding is consistent with the emerging picture from spectropolarimetry studies that CC SNe are intrinsically aspherical (see Wang \& Wheeler 2008 for a review), whereas Type Ia SNe are largely spherical (with some variation; see Kasen et al. 2009). Further observations and analyses are necessary to elucidate whether the asymmetries in CC SNRs are properties determined by the central engine/ignition process. 

As one possible end of the stellar life cycle, SNe provide crucial tests of stellar evolution models. However, in order to compare observation with theory, we need viable progenitor diagnostics for each SN type. During and immediately after explosions, many aspects of the physical mechanism can be explored: e.g., neutrinos, polarization, radioactive decays, relic compact objects. As presented here, the distribution of line emission in young SNRs gives a complementary and unique means to probe the explosion mechanism many years after a SN event. This tantalizing possibility has always been the magnet of young SNR studies, and the superb spatial and spectral resolution of {\it Chandra} has enabled detailed analyses of the metal-rich ejecta in SNRs. The clean, quantitative method we introduce here is based entirely on observables, and it represents the first standardized approach to type young SNRs. 

\vspace{5 mm}

We thank the referee, Patrick Slane, for his helpful feedback. This work is supported by DOE SciDAC DE-FC02-01ER41176 (LAL and ER-R) and a National Science Foundation Graduate Research Fellowship (LAL).

\end{document}